# Real Time Full-Color Imaging in a Meta-Optical Fiber Endoscope

*Johannes E. Fröch[1,2,\*], Luocheng Huang[2], Quentin A.A. Tanguy[2], Shane Colburn[3], Alan Zhan[3], Andrea Ravagli[4], Eric J. Seibel[5], Karl Böhringer[2,6,7], Arka Majumdar[1,2,\*]*

1: Department of Physics, University of Washington, Seattle, 98195, WA, USA

2: Department of Electrical and Computer Engineering, University of Washington, Seattle, 98195, WA, USA

3: Tunoptix, 4000 Mason Road 300, Fluke Hall, Seattle, WA, 98195 USA

4: SCHOTT North America, Inc. Southbridge, MA, 01550 USA

5: Department of Mechanical Engineering, University of Washington, Seattle, WA, 98195, USA

6: Department of Bioengineering, University of Washington, Seattle, WA, 98195, USA

7: Institute for Nano-Engineered Systems, University of Washington, Seattle, WA, 98195, USA

**Abstract**

Endoscopes are an important component for the development of minimally invasive surgeries. Their size is one of the most critical aspects, because smaller and less rigid endoscopes enable higher agility, facilitate larger accessibility, and induce less stress on the surrounding tissue. In all existing endoscopes, the size of the optics poses a major limitation in miniaturization of the imaging system. Not only is making small optics difficult, but their performance also degrades with downscaling. Meta-optics have recently emerged as a promising candidate to drastically miniaturize optics while achieving similar functionalities with significantly reduced size. Herein, we report an inverse-designed meta-optic, which combined with a coherent fiber bundle enables a 33% reduction in the rigid tip length over traditional gradient-index (GRIN) lenses. We use the meta-optic fiber endoscope (MOFIE) to demonstrate real-time video capture in full visible color, the spatial resolution of which is primarily limited by the fiber itself. Our work shows the potential of meta-optics for integration and miniaturization of biomedical devices towards minimally invasive surgery.

**Keywords:** endoscopy, meta-optics, nano-fabrication, biomedical engineering,

1. **Introduction**

Ultra-compact, agile endoscopes with large field of view (FoV), long depth of field (DoF), and short rigid tip length are important tools for the development of minimally invasive operations and new experimental surgeries.(*1–6*) As these fields develop, the requirement on miniaturization and increased precision become progressively demanding. In existing endoscopes, a fundamental limitation of the device agility within small tortuous ducts is the rigid tip length, which in turn is primarily constrained by the size of the optical elements required for imaging. Thus, alternative solutions are urgently needed to reduce the tip length.(*1, 7–10*)

An emerging and versatile idea in the photonics community to create miniaturized optical elements is meta-optics.(*11, 12*) These are sub-wavelength diffractive optical elements, composed of sub-wavelength scatterer arrays, designed to shape the phase, amplitude, and spectral response of an incident wavefront. Such ultrathin flat optics not only shrink the size of traditional optics but can also combine multiple functionalities in a single surface.(*13–18*) Hence several approaches have already explored combining meta-optics with single-mode fibers.(*19, 20*) Unfortunately, meta-optics traditionally suffer from strong aberrations (both chromatic and Seidel) making large FoV and full-color imaging challenging. Several works have shown that the common metalens design, a hyperboloid phase mask, is not suitable for simultaneously capturing color information across the visible spectrum, typically resulting in images, that are crisp for the design wavelength (e.g. green) but strongly aberrated/ blurred for other colors (red, and blue).(*21–23*) While dispersion engineering,(*24, 25, 25–27*) and computational imaging techniques(*13, 21, 23*) can reduce chromatic aberration, they either suffer from small apertures, low numerical aperture,(*28*, *29*) or require a computational post-processing step, complicating real-time video capture. Similarly, an additional aperture in front of the meta-optic can provide a larger FoV,(*30*, *31*) but comes at the cost of reduced light collection and increased thickness of the optics.

In this work, we demonstrate an inverse-designed meta-optic, optimized to capture real-time full-color scenes in the visible in conjunction with a 1 mm diameter coherent fiber bundle (Figure 1). The meta-optic enables operations at a FoV of 30$^o$, a DoF of > 30 mm (exceeding 300 % of the nominal design working distance) and a minimum rigid tip length of only ~ 2.5 mm. Compared to a traditional gradient-index (GRIN) lens integrated fiber bundle endoscope, this is a 33% tip length reduction, thanks to the shorter focal length and the ultrathin nature of the meta-optic, while at the same time comparable imaging performance and working distance is maintained. To achieve exceptional FoV, DoF, and color performance of the Meta-Optical Fiber Endoscope (MOFIE), we approached this design problem from a system level perspective, considering that the diameter and spacing of individual fiber cores within the bundle limit the achievable image quality, which in turn also limits the achievable FoV(*6*). This aspect is implemented in an automatic differentiation framework using the volume of a multichromatic, modulation transfer function (MTF) as the figure of merit, while a phase mask is iteratively adapted to maximize that figure of merit. The design algorithm is further detailed in the supplementary information. By ensuring the meta-optic has a MTF within the limit posed by the fiber bundle, we achieve full color operation without the requirement of a computational reconstruction step, thus facilitating real time operation.

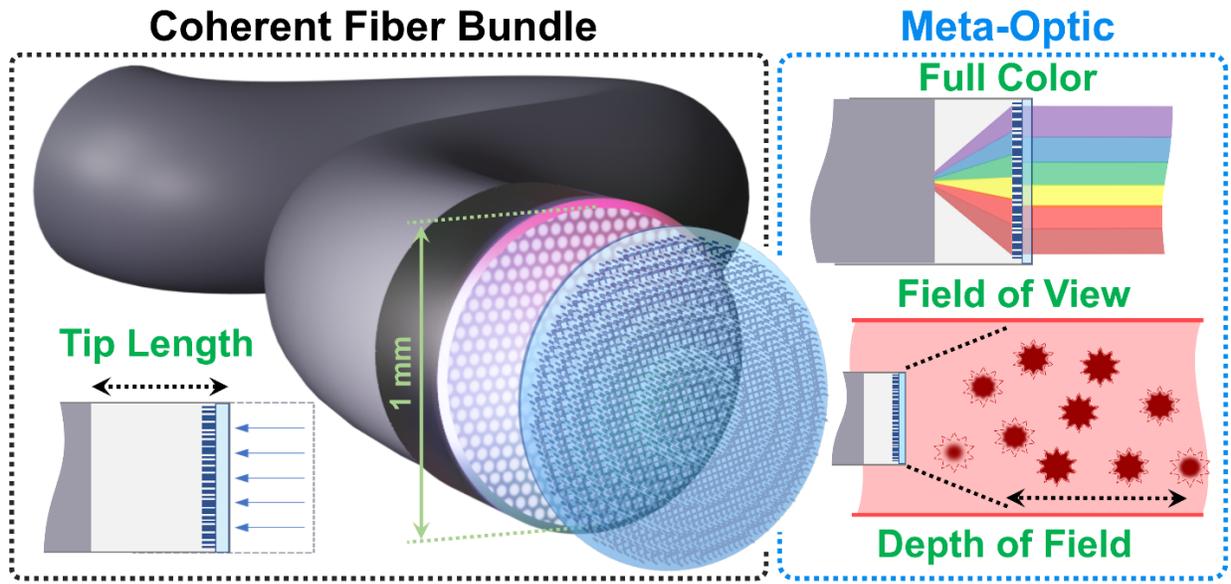

*Figure 1. Schematic of the meta-optical fiber endoscope: the meta-optics is optimized to have a large average volume under the MTF curve over a broad wavelength range to allow color preservation. In comparison with a traditional GRIN lens, the use of meta-optics reduces the tip length, while maintaining a wide field of view of $30°$ and large depth of field exceeding 30 mm.*

2. Results

A defining limitation of meta-optics is their chromatic aberration.(*32*) Whereas a hyperboloid meta-lens can have diffraction-limited performance at the design wavelength, the MTF degrades rapidly for operation over an extended spectral range.(*29*) To tackle this issue, we use an inverse design approach, which maximizes the volume under a multi-chromatic MTF curve as the figure of merit during the optimization. Our figure of merit also ensures that the MTF remains similar over a broad wavelength range, which is critical for broadband operation (detailed in the Supplementary Information). To ensure polarization-insensitive operation and compatibility with high volume manufacturing processes, the specific design is implemented through simple square pillars with a minimum feature size of 75 nm and a maximum aspect ratio of 10 (detailed in the Supplementary Information).

We fabricated the 1 mm aperture f/2 meta-optic in SiN, due to its high transparency in the visible wavelength range (details in Methods) and wide availability of thin films grown by plasma-enhanced chemical vapor deposition on quartz.(33) An optical image (Figure 2a) shows that the device integrity is maintained throughout the entire area, exhibited by the same structural color for specific radii of the meta-optic stemming from a radial-symmetric design constraint. Further inspection with a scanning electron microscope of the meta-optic (Figure 2b,c) shows the scatterer quality of a representative device area, highlighting the successful fabrication of the individual elements with the desired footprint and aspect ratio, as well as negligible sidewall roughness (Figure 2d). Detailed characterization, such as the point spread function (PSF) and the MTF of the stand- alone meta-optic can be found in the Supplementary Information.

For a demonstration of the MOFIE, we placed an OLED screen in front of the meta-optic at the proximal end of the CFB and a capturing system at its distal end, consisting of an objective, tube lens, and camera. The setup is further detailed in the Methods section and schematically depicted in the Supplementary Information. A set of multicolored images were then displayed on the OLED screen (Figure 2e, f) and captured through the MOFIE as shown in Figure 2g, h. Importantly, the color quality is entirely preserved throughout the visible range, even for more complex scenes (Figure 2f, h), while maintaining a reasonable resolution. This is relevant, because common metalens designs (such as a hyperboloid) are well known to have a very strong chromatic aberration, which precludes full-color operation over the entire visible wavelength range. Yet, medical surgery requires high quality color reproduction to confidently discern diseased tissue.

We emphasize here that in comparison to many prior works(*13, 21, 22, 34*), the presented images were not computationally deconvolved to enhance the image quality.

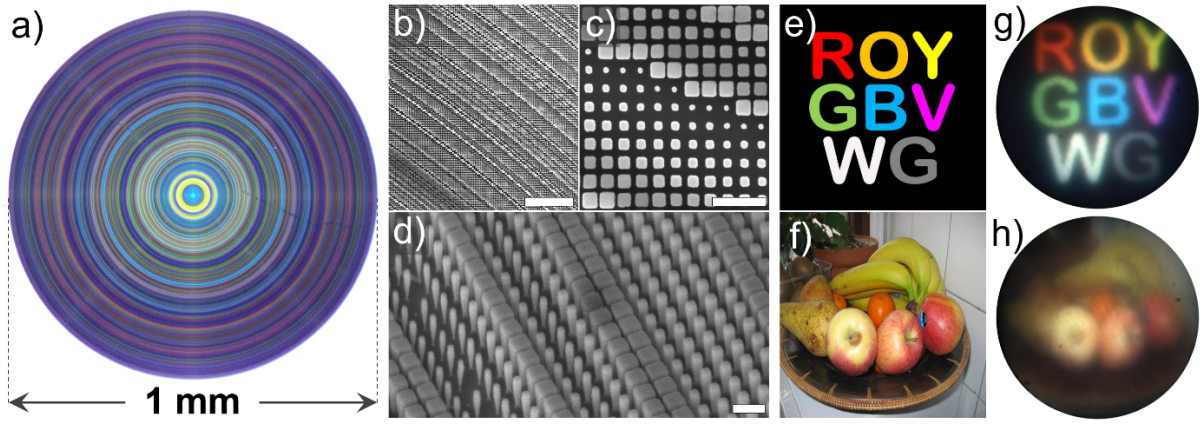

*Figure 2. Characterization of the meta-optic. a) Optical microscope image of the 1 mm aperture meta-optic. Different colors of the device correspond to different regions of pillars, which exhibit a structural color effect. b), c) Top view SEM images of the fabricated device. Scale bars correspond to 10 µm and 1 µm, respectively. d) SEM image at an oblique angle of 45°. The scale bar corresponds to 500 nm. e), f) Full color scenes displayed on an OLED screen. g), h) Corresponding images captured with the MOFIE at working distance of 10 mm. Images displayed are without any computational deconvolution.*

In the following we characterized the performance of the MOFIE, with respect to its FoV, DoF, and spatial resolution using the previously described imaging configuration. The results are summarized in Figure 3.

First, from a measurement of a checkerboard pattern (Figure 3a, b), at a working distance of 10 mm, we determined the FoV to be ~ 30°. In context, this still allows to view an object area with a field of view of ~ 5 mm at a working distance of 10 mm. This is relevant because the average diameter of several important human arteries (e.g. coronary, cerebral) are on the order of ~ 1 - 3 mm.(*35*, *36*) Hence the MO endoscope completely covers the entire relevant cross section that will be under investigation during operation.

A further important factor for endoscope imaging is the depth of field, because *in-vivo* movements such as a beating heart or breathing lung, can significantly change the working distance within tens of milliseconds, making a large DoF extremely beneficial. To evaluate the performance in this regard, we placed the OLED screen at different working distances of the respective endoscope, while displaying the same images, at working distances of 7 mm, 10 mm (design working distance), 18 mm, and 40 mm (intermediates are shown in the Supplementary Information). This was done while maintaining the same distance between the coherent fiber bundle and meta-optic. Importantly, it can be clearly seen from the image series (Figure 3c-e) that throughout this range of 33 mm the same sharpness and color quality is achieved.

To quantitatively assess the imaging resolution at the design working distance (10 mm), we placed a USAF Target in front of the MOFIE, which was illuminated by a broadband halogen source from a second fiber placed next to the CFB. A captured image of the group 3 elements is shown in Figure 3f. A magnified image in 3g, shows the attainable resolution down to elements 3 and 4, indicating a line resolution of ~ 50 µm. An important aspect is that the individually resolved color channels show the same resolution throughout, indicated by similar lines profiles. We note that the yellowish tint in the captured images stems from the spectral distribution of the broad band halogen source, which however can be rebalanced on the image display or resolved using a spectrally equal distributed light source.

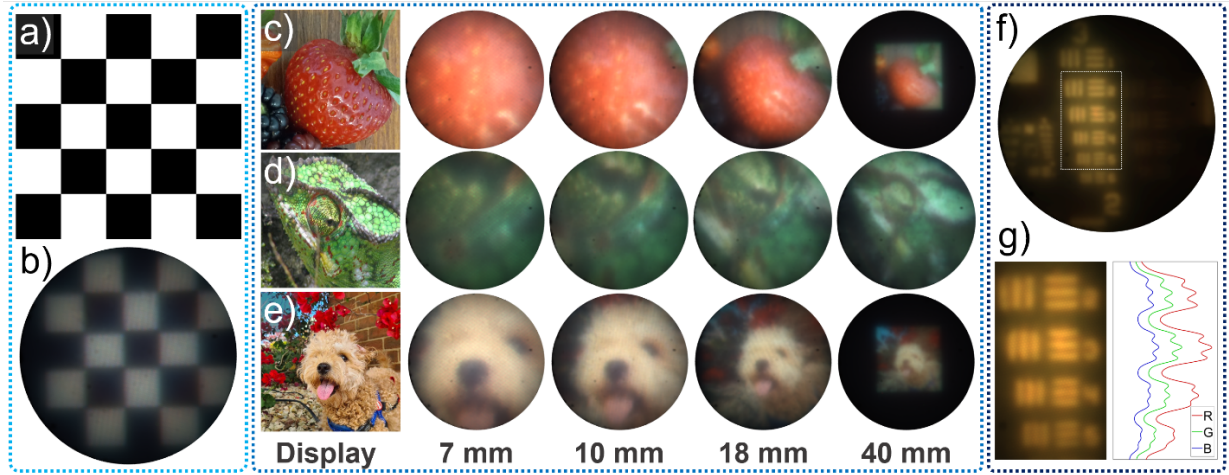

*Figure 3. Assessment of the MOFIE. a) A checkerboard pattern as displayed on a screen with a width of 4.5 mm. b) Captured image through the MOFIE at a working distance of 10 mm. c), d), and e), show images that were displayed on an OLED screen and captured at working distances of 7 mm, 10 mm, 18 mm, and 40 mm. The size of images c) and e) on the display was 4.5 mm. Image d) was displayed with a size of 8 mm. f) Captured image of the group 3 lines of a USAF 1951 resolution chart. g) shows a magnified image of elements 2 – 5, with a line profile of the red (R), green (G), and blue (B) channels.*

To emphasize the capabilities of the MOFIE we demonstrate real-time, full-color imaging of a biological sample, shown in Figure 4. This is an important step, because in-the-field deployment of an endoscope ultimately requires imaging capabilities with video-rate speed while maintaining full-color information. To emulate such a situation under the given requirements, we recorded the movement of a living caterpillar on top of a strawberry leaf at a video rate of ~ 14 frames-per-second at a working distance of ~ 10 mm (a picture of the image scene is shown in the Supplementary Information). For illumination of the scene, we used a broadband tungsten-halogen source, which was coupled into an optical fiber, placed next to the coherent fiber bundle pointing towards the scene.

In Figure 4, a frame series is shown as the caterpillar moves across the leaf, the full movie can be found in the Supplementary Information. We note that no deconvolution of the images was applied, and the shown frames are equivalent to the imaging that was displayed on the image capture tool during this experiment. We emphasize here that this recording is in fact to our knowledge, one of the first demonstrations of full-color video-rate imaging using a meta-optic, which further highlights a relevant advancement for meta-optics.

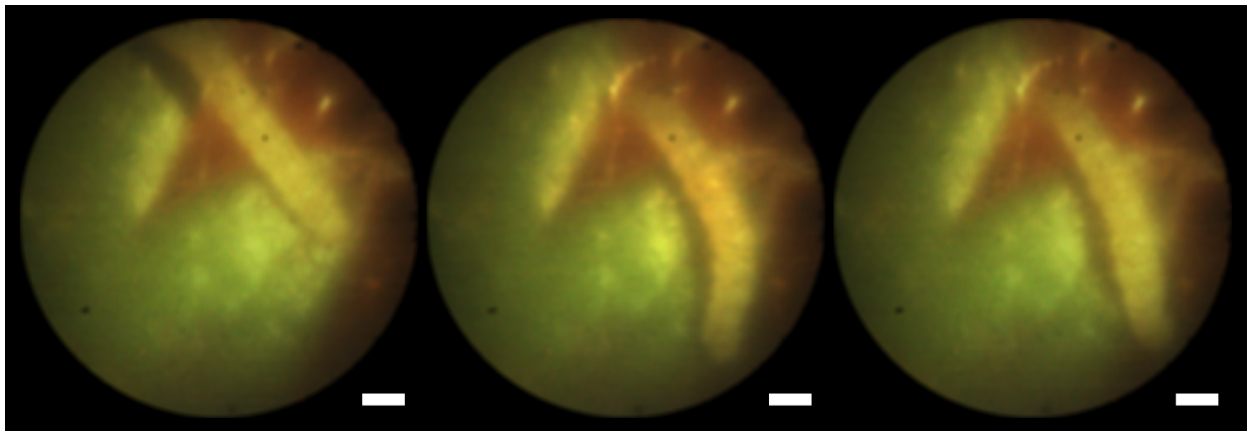

*Figure 4. Clip of a full color video rate imaging of a caterpillar on a strawberry leaf. Observation of a crawling movement of the caterpillar across a gap in the strawberry leaf. White scale bars correspond to 500 μm. A yellowish tint on the captured video appeared due to the spectral distribution of the used light source (a tungsten halogen lamp).*

3. Discussion

To assess the performance metrics of the MOFIE relative to an established system, we further compared it with a graded index (GRIN) lens-based endoscope. A comparison of images that were displayed on the OLED screen and images that were captured either with the MOFIE or the GRIN endoscope are summarized in the Supplementary Information.

We reduced the optical tip length from 3.75 mm for a GRIN lens-based endoscope by 33% down to 2.5 mm, for the MOFIE. We note that the tip length in MOFIE can be readily reduced further by using a thinner carrier substrate (100 μm instead of 500 μm). The FoV of the MOFIE is 30° compared to 45° for the GRIN lens-based endoscope. In contrast, we achieve a better DoF perception over the characterized working distance range (~ 30 mm). While we used a benchtop configuration in the current demonstration, future effort will focus on full integration with a coherent fiber bundle and direct medical demonstrations to illustrate the full power of the MOFIE. Beyond that, multiple pathways could be pursued integrating meta-optics with coherent fiber bundles. Specifically, a suitable computational backend might further improve the image quality or arrayed meta-optics(*37*) could enable an ultra large field of view or foveated imaging.

4. **Conclusion**

In summary, we have demonstrated real-time, full-color operation of a meta-optic fiber endoscope (MOFIE). Approaching this problem from a system level design perspective, we created a meta-optic that is specifically suited for operation in combination with the optical constraints of a coherent fiber bundle. Specifically, as the spatial resolution of MOFIE is limited by the coherent fiber bundle, the meta-optic is optimized to have an MTF sufficient to image through the fiber bundle. Thus, the combined system facilitates full-color, video-rate imaging, which could be potentially further improved using a computational backend. However, in this case a considerable optimized algorithm computer hardware would be required to avoid any lag during operation of the MOFIE.

The fabricated meta-optic has a device length of 2.5 mm, which is a reduction of ~ 33% to a comparable GRIN lens (3.75 mm). We demonstrate a 30° field of view, >30 mm depth of field, and real time, full color imaging, highlighted by video recordings of a biological sample. To the best of our knowledge the present work presents one of the first applications, where a meta-optic is used for full color and real time imaging simultaneously. Beyond that, our work exemplifies where meta-optics can potentially find a strong position in biomedical applications with drastically reduced imaging system size.

5. Methods

**Fabrication**

For fabrication, we first deposited a ~ 750 nm thick SiN film on a 500 µm thick quartz wafer using plasma enhanced chemical vapor deposition (PECVD) in a SPTS PECVD chamber. A positive resist (ZEP 520A) was then spun onto the wafer, followed by baking at 180 °C for 3 minutes. To minimize charging effects during patterning, a conductive polymer layer (DisCharge H2O) was subsequently spun on top. The resist layer was then patterned using a 100 kV electron beam (JEOL JBX6300FS) at a dose of ~ 300 µC cm$^{-2}$ and developed in Amyl Acetate for 2 minutes. Then a layer (~ 50 nm) of AlO$_x$ was deposited using electron beam evaporation. After overnight liftoff in NMP, the SiN layer was etched using a mixture of $C_4F_8$/$SF_6$ in an inductively coupled reactive ion etcher (Oxford PlasmaLab System 100). For SEM imaging a thin conductive Au/Pd layer was deposited.

**Measurements**

The measurement setup consisted of an OLED screen, the meta-optic element andthe CFB (SCHOTT RLIB CVET, 1.05X 910, 7.6M, 18K19, QA.90). The imaging setup at the distal end of the CFB consisted of an objective lens (Nikon), a tube lens (Thorlabs), and a camera (Allied Vision GT 1930 C). In configurations where the USAF 1951 target was imaged, or for video recording, the sample was illuminated with a broadband halogen source (Thorlabs SLS301), which was delivered and directed towards the sample through a multimode fiber (core size 200 µm). The fiber was placed next to the CFB, which would be a typical configuration for an endoscope. For comparison with a GRIN lens endoscope, a GRIN lens (GoFoton) with 1.3 mm lens diameter, length of 3.75 mm, and working distance of 10 mm was attached to the CFB.

During processing of the acquired images, we have only adjusted the brightness, as well as adding a slight blur to the image to minimize artificial moiré patterns that would otherwise emerge due to the periodic arrangement of the fiber cores in the CFB. Such a slight deblur, however, could also be achieved by adjusting the output facet of the fiber relative to the imaging system. A comparison is shown in the Supplementary Information.

**Acknowledgements**

This research was supported by NSF-GCR-2120774. Part of this work was conducted at the Washington Nanofabrication Facility / Molecular Analysis Facility, a National Nanotechnology Coordinated Infrastructure (NNCI) site at the University of Washington with partial support from the National Science Foundation via awards NNCI-1542101 and NNCI-2025489.

**Conflict of Interest**

AZ, SC, KB and AM are part of Tunoptix, a company commercializing this technology.

**References**
1. B. A. Flusberg, E. D. Cocker, W. Piyawattanametha, J. C. Jung, E. L. M. Cheung, M. J. Schnitzer, Fiber-optic fluorescence imaging. *Nat Methods*. **2**, 941–950 (2005).
2. C. M. Lee, C. J. Engelbrecht, T. D. Soper, F. Helmchen, E. J. Seibel, Scanning fiber endoscopy with highly flexible, 1 mm catheterscopes for wide-field, full-color imaging. *Journal of Biophotonics*. **3**, 385–407 (2010).
3. A. Perperidis, K. Dhaliwal, S. McLaughlin, T. Vercauteren, Image computing for fibre-bundle endomicroscopy: A review. *Medical Image Analysis*. **62**, 101620 (2020).
4. A. Shadfan, A. Hellebust, R. Richards-Kortum, T. Tkaczyk, Confocal foveated endomicroscope for the detection of esophageal carcinoma. *Biomed. Opt. Express, BOE*. **6**, 2311–2324 (2015).
5. A. Shadfan, H. Darwiche, J. Blanco, A. Gillenwater, R. Richards-Kortum, T. S. Tkaczyk, Development of a multimodal foveated endomicroscope for the detection of oral cancer. *Biomed. Opt. Express, BOE*. **8**, 1525–1535 (2017).

6. J. Park, D. J. Brady, G. Zheng, L. Tian, L. Gao, Review of bio-optical imaging systems with a high space-bandwidth product. *AP*. **3**, 044001 (2021).
7. J. Shin, D. N. Tran, J. R. Stroud, S. Chin, T. D. Tran, M. A. Foster, A minimally invasive lens-free computational microendoscope. *Science Advances*. **5**, eaaw5595 (2019).
8. N. Badt, O. Katz, Real-time holographic lensless micro-endoscopy through flexible fibers via fiber bundle distal holography. *Nat Commun*. **13**, 6055 (2022).
9. A. Orth, M. Ploschner, E. R. Wilson, I. S. Maksymov, B. C. Gibson, Optical fiber bundles: Ultra-slim light field imaging probes. *Science Advances*. **5**, eaav1555 (2019).
10. W. Choi, M. Kang, J. H. Hong, O. Katz, B. Lee, G. H. Kim, Y. Choi, W. Choi, Flexible-type ultrathin holographic endoscope for microscopic imaging of unstained biological tissues. *Nat Commun*. **13**, 4469 (2022).
11. N. Yu, F. Capasso, Flat optics with designer metasurfaces. *Nature Mater*. **13**, 139–150 (2014).
12. M. K. Chen, Y. Wu, L. Feng, Q. Fan, M. Lu, T. Xu, D. P. Tsai, Principles, Functions, and Applications of Optical Meta-Lens. *Advanced Optical Materials*. **9**, 2001414 (2021).
13. S. Colburn, A. Zhan, A. Majumdar, Metasurface optics for full-color computational imaging. *Science Advances*. **4**, eaar2114 (2018).
14. C. Munley, W. Ma, J. E. Fröch, Q. A. A. Tanguy, E. Bayati, K. F. Böhringer, Z. Lin, R. Pestourie, S. G. Johnson, A. Majumdar, Inverse-Designed Meta-Optics with Spectral-Spatial Engineered Response to Mimic Color Perception. *Advanced Optical Materials*. **10**, 2200734 (2022).
15. A. Arbabi, Y. Horie, M. Bagheri, A. Faraon, Dielectric metasurfaces for complete control of phase and polarization with subwavelength spatial resolution and high transmission. *Nature Nanotech*. **10**, 937–943 (2015).
16. M. Piccardo, V. Ginis, A. Forbes, S. Mahler, A. A. Friesem, N. Davidson, H. Ren, A. H. Dorrah, F. Capasso, F. T. Dullo, B. S. Ahluwalia, A. Ambrosio, S. Gigan, N. Treps, M. Hiekkamäki, R. Fickler, M. Kues, D. Moss, R. Morandotti, J. Riemensberger, T. J. Kippenberg, J. Faist, G. Scalari, N. Picqué, T. W. Hänsch, G. Cerullo, C. Manzoni, L. A. Lugiato, M. Brambilla, L. Columbo, A. Gatti, F. Prati, A. Shiri, A. F. Abouraddy, A. Alù, E. Galiffi, J. B. Pendry, P. A. Huidobro, Roadmap on multimode light shaping. *J. Opt.* **24**, 013001 (2021).
17. S. C. Malek, A. C. Overvig, A. Alù, N. Yu, Multifunctional resonant wavefront-shaping meta-optics based on multilayer and multi-perturbation nonlocal metasurfaces. *Light Sci Appl*. **11**, 246 (2022).
18. Z. Lin, R. Pestourie, C. Roques-Carmes, Z. Li, F. Capasso, M. Soljačić, M. Soljačić, S. G. Johnson, End-to-end metasurface inverse design for single-shot multi-channel imaging. *Opt. Express, OE*. **30**, 28358–28370 (2022).
19. H. Ren, J. Jang, C. Li, A. Aigner, M. Plidschun, J. Kim, J. Rho, M. A. Schmidt, S. A. Maier, An achromatic metafiber for focusing and imaging across the entire telecommunication range. *Nat Commun*. **13**, 4183 (2022).
20. H. Pahlevaninezhad, M. Khorasaninejad, Y.-W. Huang, Z. Shi, L. P. Hariri, D. C. Adams, V. Ding, A. Zhu, C.-W. Qiu, F. Capasso, M. J. Suter, Nano-optic endoscope for high-resolution optical coherence tomography in vivo. *Nature Photon*. **12**, 540–547 (2018).
21. E. Tseng, S. Colburn, J. Whitehead, L. Huang, S.-H. Baek, A. Majumdar, F. Heide, Neural nano-optics for high-quality thin lens imaging. *Nat Commun*. **12**, 6493 (2021).

22. L. Huang, J. Whitehead, S. Colburn, A. Majumdar, A. Majumdar, Design and analysis of extended depth of focus metalenses for achromatic computational imaging. *Photon. Res., PRJ*. **8**, 1613–1623 (2020).
23. L. Huang, S. Colburn, A. Zhan, A. Majumdar, Full-Color Metaoptical Imaging in Visible Light. *Advanced Photonics Research*. **n/a**, 2100265.
24. W. T. Chen, A. Y. Zhu, V. Sanjeev, M. Khorasaninejad, Z. Shi, E. Lee, F. Capasso, A broadband achromatic metalens for focusing and imaging in the visible. *Nature Nanotech*. **13**, 220–226 (2018).
25. S. Wang, P. C. Wu, V.-C. Su, Y.-C. Lai, M.-K. Chen, H. Y. Kuo, B. H. Chen, Y. H. Chen, T.-T. Huang, J.-H. Wang, R.-M. Lin, C.-H. Kuan, T. Li, Z. Wang, S. Zhu, D. P. Tsai, A broadband achromatic metalens in the visible. *Nature Nanotech*. **13**, 227–232 (2018).
26. W. T. Chen, A. Y. Zhu, F. Capasso, Flat optics with dispersion-engineered metasurfaces. *Nat Rev Mater*. **5**, 604–620 (2020).
27. W. Zang, Q. Yuan, R. Chen, L. Li, T. Li, X. Zou, G. Zheng, Z. Chen, S. Wang, Z. Wang, S. Zhu, Chromatic Dispersion Manipulation Based on Metalenses. *Advanced Materials*. **32**, 1904935 (2020).
28. F. Presutti, F. Monticone, Focusing on bandwidth: achromatic metalens limits. *Optica, OPTICA*. **7**, 624–631 (2020).
29. J. Engelberg, U. Levy, Achromatic flat lens performance limits. *Optica, OPTICA*. **8**, 834–845 (2021).
30. M. Y. Shalaginov, S. An, F. Yang, P. Su, D. Lyzwa, A. M. Agarwal, H. Zhang, J. Hu, T. Gu, Single-Element Diffraction-Limited Fisheye Metalens. *Nano Lett.* **20**, 7429–7437 (2020).
31. J. Engelberg, C. Zhou, N. Mazurski, J. Bar-David, A. Kristensen, U. Levy, Near-IR wide-field-of-view Huygens metalens for outdoor imaging applications. *Nanophotonics*. **9**, 361–370 (2020).
32. E. Arbabi, A. Arbabi, S. M. Kamali, Y. Horie, A. Faraon, Multiwavelength polarization-insensitive lenses based on dielectric metasurfaces with meta-molecules. *Optica, OPTICA*. **3**, 628–633 (2016).
33. S. Colburn, A. Zhan, E. Bayati, J. Whitehead, A. Ryou, L. Huang, A. Majumdar, Broadband transparent and CMOS-compatible flat optics with silicon nitride metasurfaces [Invited]. *Opt. Mater. Express, OME*. **8**, 2330–2344 (2018).
34. E. Bayati, R. Pestourie, S. Colburn, Z. Lin, S. G. Johnson, A. Majumdar, Inverse designed extended depth of focus meta-optics for broadband imaging in the visible. *Nanophotonics* (2021), doi:10.1515/nanoph-2021-0431.
35. P. Mouches, N. D. Forkert, A statistical atlas of cerebral arteries generated using multi-center MRA datasets from healthy subjects. *Sci Data*. **6**, 29 (2019).
36. J. T. Dodge Jr, B. G. Brown, E. L. Bolson, H. T. Dodge, Lumen diameter of normal human coronary arteries. Influence of age, sex, anatomic variation, and left ventricular hypertrophy or dilation. *Circulation*. **86**, 232–246 (1992).
37. J. Chen, J. Chen, X. Ye, X. Ye, S. Gao, S. Gao, Y. Chen, Y. Chen, Y. Zhao, Y. Zhao, C. Huang, K. Qiu, S. Zhu, S. Zhu, T. Li, T. Li, Planar wide-angle-imaging camera enabled by metalens array. *Optica, OPTICA*. **9**, 431–437 (2022).

# Supplementary Information:

# Real Time Full-Color Imaging in a Meta-Optical Fiber Endoscope

*Johannes E. Fröch[1,2,*], Luocheng Huang[2], Quentin A.A. Tanguy[2], Shane Colburn[3], Alan Zhan[3], Andrea Ravagli[4], Eric J. Seibel[5], Karl Böhringer[2,6,7], Arka Majumdar[1,2,*]*

1: Department of Physics, University of Washington, Seattle, 98195, WA, USA

2: Department of Electrical and Computer Engineering, University of Washington, Seattle, 98195, WA, USA

3: Tunoptix, 4000 Mason Road 300, Fluke Hall, Seattle, WA, 98195 USA

4: SCHOTT North America, Inc. Southbridge, MA, 01550 USA

5: Department of Mechanical Engineering, University of Washington, Seattle, WA, 98195, USA

6: Department of Bioengineering, University of Washington, Seattle, WA, 98195, USA

7: Institute for Nano-Engineered Systems, University of Washington, Seattle, WA, 98195, USA

## Content

**S.1 Scatterer**
**S.2 Design**
**S.3 PSF and MTF**
**S.4 Measurement Setup**
**S.5 Additional Characterization**
**S.6 Image Scene of Video**
**S.7 Comparison to a GRIN lens Endoscope**
**S.8 Additional Image Processing Comparison**

## S.1 Scatterer

To identify suitable scatterers to implement the meta-optic we used rigorous coupled wave analysis (RCWA) to calculate the transmission and phase response for square pillars in SiN on a quartz substrate, considering height, period, and width as variable parameters. From these calculations we chose scatterers with varying width (*a*), period of 350 nm, and height of 750 nm on top of 100 nm SiN film on top of a quartz substrate (~ 500 µm). By adjusting the pillar side width, *a*, a full 0 - 2π phase shift of the transmitted light can be achieved. Exemplary scatterer responses for wavelengths of 460 nm, 540 nm, and 620 nm are shown in Figure S1, respectively.

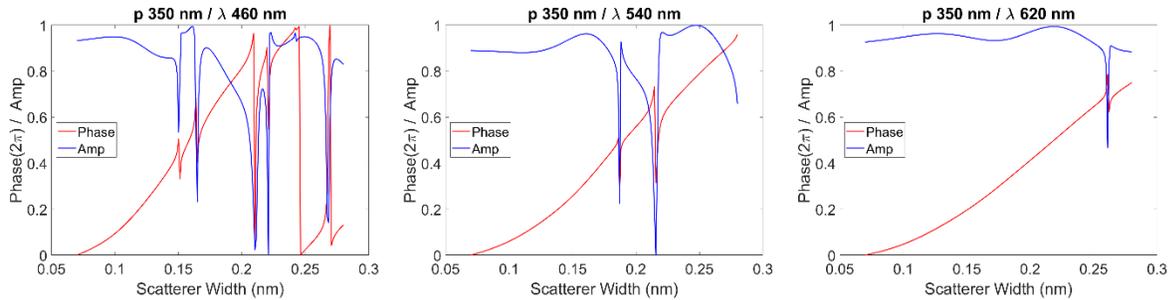

*Figure S1.a) Schematic of the scatterer geometry. b) Phase and amplitude response for wavelengths of 460 nm, 560 nm, and 660 nm, respectively.*

## S.2 Design

Hyperboloid meta-lenses are known to have very strong chromatic aberration, because the phase wrapping point of $2\pi$ is reached at different locations for different wavelengths. Specifically, the focal length is inversely proportional to the wavelength of the light [1]. Thus, if we find a sensor plane where one wavelength (e.g. green) is well-focused, the other wavelengths are defocused. Hence, the modulation transfer function (MTF) of the defocused wavelength is very narrow, and thus fails to capture enough information that can be inverted to reconstruct the image. In contrast, in a true-color imaging scenario, the input light is broadband, i.e. not only composed of three wavelengths, but a continuous spectrum. Given that the color-filters at the detector side are also broad-band, we need to ensure that we preserve a good enough MTF for all the color bands over the entire wavelength range. Thus, to have full-color imaging, we need to ensure that the MTF remains same over the whole range of bandwidth, while simultaneously having large volume under the MTF curve.

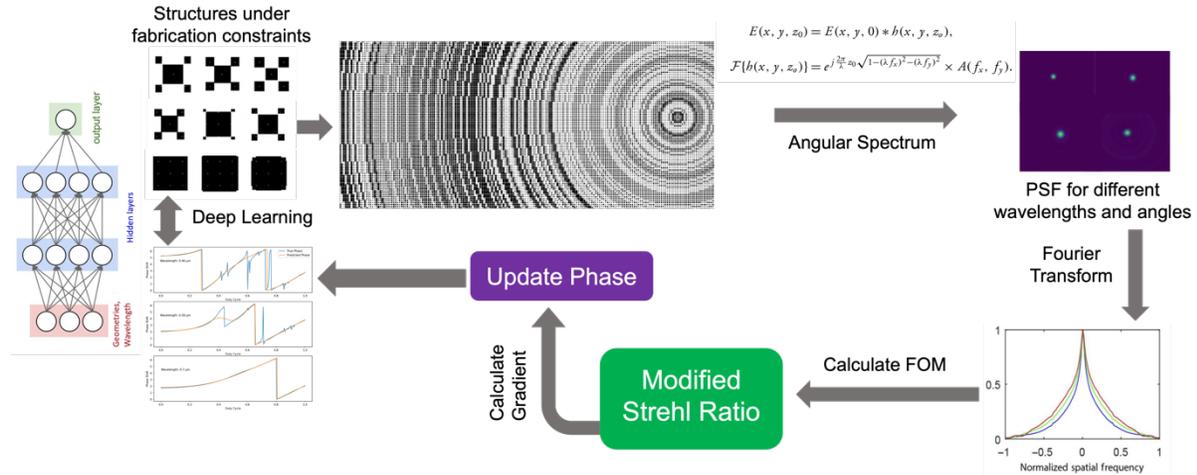

*Figure S2. Design flow for MTF-engineered meta-optics for full-color imaging.*

We used an automatic differentiation framework to optimize the meta-optics. We first created a deep learning mapping between scatterer geometry and phase. Such a mapping is differentiable, and we can directly optimize the scatterers. While such a framework can be used to design complex parametrized scatterers, to ensure compatibility with high volume manufacturing we employ simple square pillar in this work. The phase-mask is propagated to the sensor plane via band-limited angular spectrum method for different wavelengths ($\lambda$) and incidence angles ($\theta$). We then calculate the MTF by taking the Fourier transform of the PSF. We then define a Strehl Ratio (SR) by the ratio of the volume under the MTF curve and volume under a diffraction limited MTF curve. Then we define the Figure of Merit (FOM) as

$$FOM = -\sum_{\lambda,\theta} \log(SR(\lambda,\theta)) = -\log\left(\prod_{\lambda,\theta} SR(\lambda,\theta)\right)$$

Such a FOM in minimized when $SR(\lambda,\theta)$ is high, while all the $SR(\lambda,\theta)$ are similar, as evident from the fact that a geometric mean is highest when all terms are similar. Ideally, we should choose the wavelength and incidence angles over the whole desired range with high resolution. However, a large number of wavelengths and incidence angles increase the optimization time. In our case, we optimized the meta-optics at 10nm wavelength increments and $5^o$ increments of incidence angle.

**S.3 PSF and MTF**

The imaging capability of the meta optic was assessed using a microscope relay setup, schematically shown in Figure S3a. The emission of an LED (either 455nm, 525nm, or 625nm) is coupled to a single mode fiber, which is placed at a distance of ~ 15 cm in front of the sample, acting as a point source. The light is then transmitted through and focused by the meta optic, whereas the resulting focal spot is captured with an objective lens and depicted on a camera through a tube lens. Images of the captured intensity profiles are shown in Figure S3b for R, G, and B, from left to right, respectively. From the corresponding intensity profiles (Figure S3b) we extracted the point spread function (PSF-Figure S3c) and calculated the modular transfer function (MTF-Figure S3d). From the PSF we observe a FWHM of about ~ 1.2 um for all three color channels. In addition, the MTF indicates similar imaging capabilities for all color channels, although the blue channels is slightly worse than others. However, as shown in the main text Figures 2 and 3, the performance is still sufficient to clearly identify true color images.

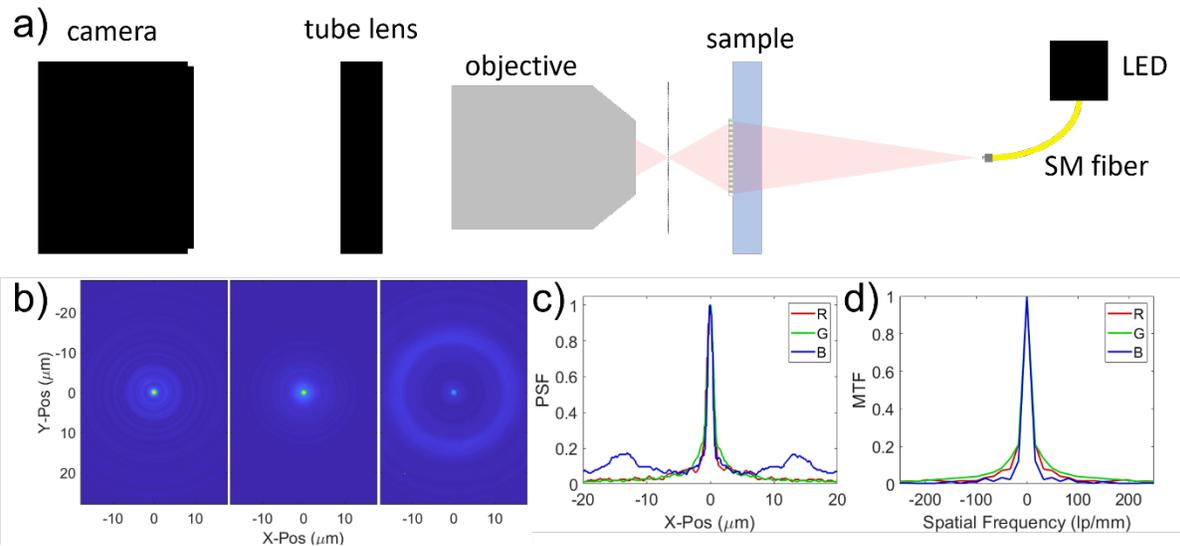

*Figure S3.a) Schematic of the measurement setup to retrieve the PSF. Elements are not too scale. b) Intensity maps of the PSFs for R, G, and B, from left to right, respectively. b) Linecuts of the PSF. c) Linecuts of the MTF.*

**S.4 Measurement Configuration**

To assess the capability of the MOFIE, we acquired various scenes in a setup as shown in Figure S4. Here the meta optic was placed in front of the distal end of the coherent fiber bundle at a distance of ~ 2 mm away from the fiber bundle. An image or scene (either an OLED display, a resolution target, or a caterpillar on a leave) was then placed in front of the MOFIE for specific measurements. The image collected and focused into the coherent fiber bundle was then retrieved at the proximal end using an objective lens. The image was then depicted on a camera (Allied Vision Pro Silica GT) through a tube lens (Thorlabs).

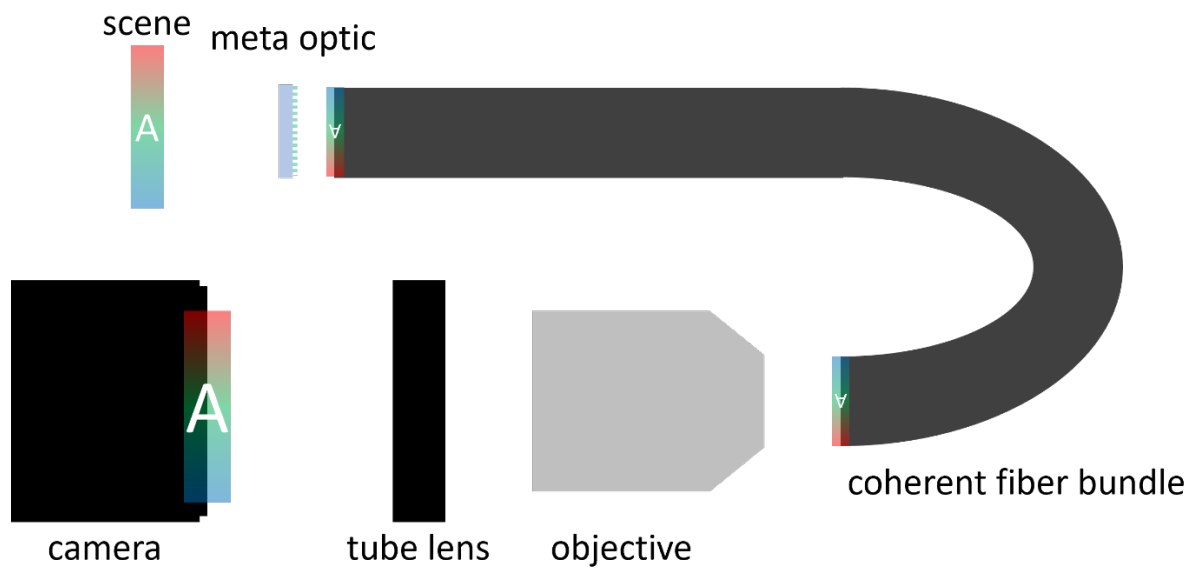

*Figure S4. Schematic of the measurement configuration for the MOFIE.*

## S5. Image Scene of the Video

To demonstrate full color and real time operation of the MOFIE, we created an imaging scene consisting of a static strawberry leaf (Figure S5a), and a living caterpillar (Figure S5b), which moved across the leaf. The scene was placed vertically in front of the MOFIE at a working distance of ~ 10 mm, and illuminated by a broad halogen light source, which was delivered to the imaging scene through a multimode fiber placed next to the coherent fiber bundle. Frames were acquired with an exposure time of ~ 70 ms.

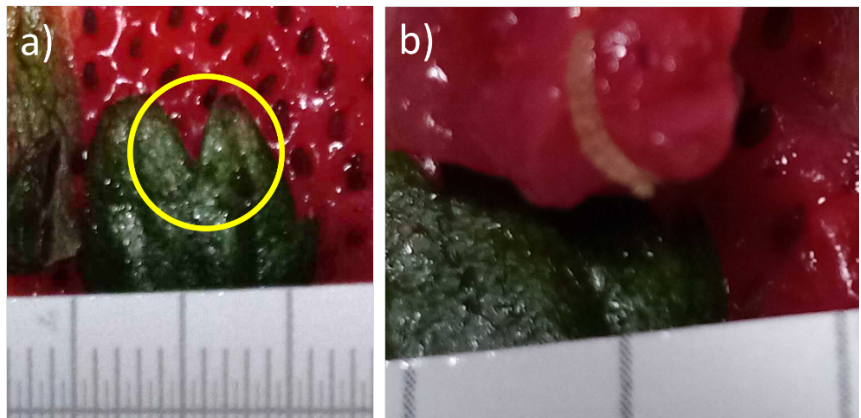

*Figure S5. Photograph of the imaging scene including a caterpillar on a strawberry leave. a) The strawberry leaf, where the caterpillar was placed upon with the imaged area highlighted by a*

*yellow circle. b) Image of the caterpillar. The ruler placed next to both objects has major tick marks of 5 mm and minor tick marks of 1 mm.*

**S.6 Comparison to a GRIN lens Endoscope and additional Characterization**

To compare the size reduction that is achievable by the MOFE, we placed it directly next to a GRIN lens based endoscope, shown in Figure S6. The overall length reduction of the optical element corresponds to ~ 1mm, which is about 30 % shorter than the GRIN lens. To reduce the overall tip length further, the thickness of the substrate (0.5 mm in our prototype) which holds the meta optic could be further reduced.

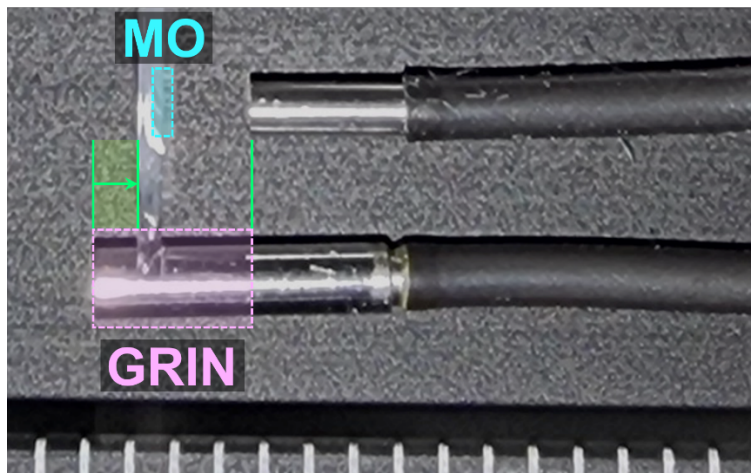

*Figure S6. Direct comparison of the tip length reduction between the meta optic (MO) and the GRIN lens directly with the coherent fiber bundle.*

Further images at intermediate working distances in addition to what has been shown in Figure 3, are shown in the top three rows in Figure S7. In addition, the bottom three rows show images taken with a GRIN lens in front of the coherent fiber bundle. Notably, for images taken at larger depth of focus, the image quality taken with GRIN degrades more than for the MOFIE.

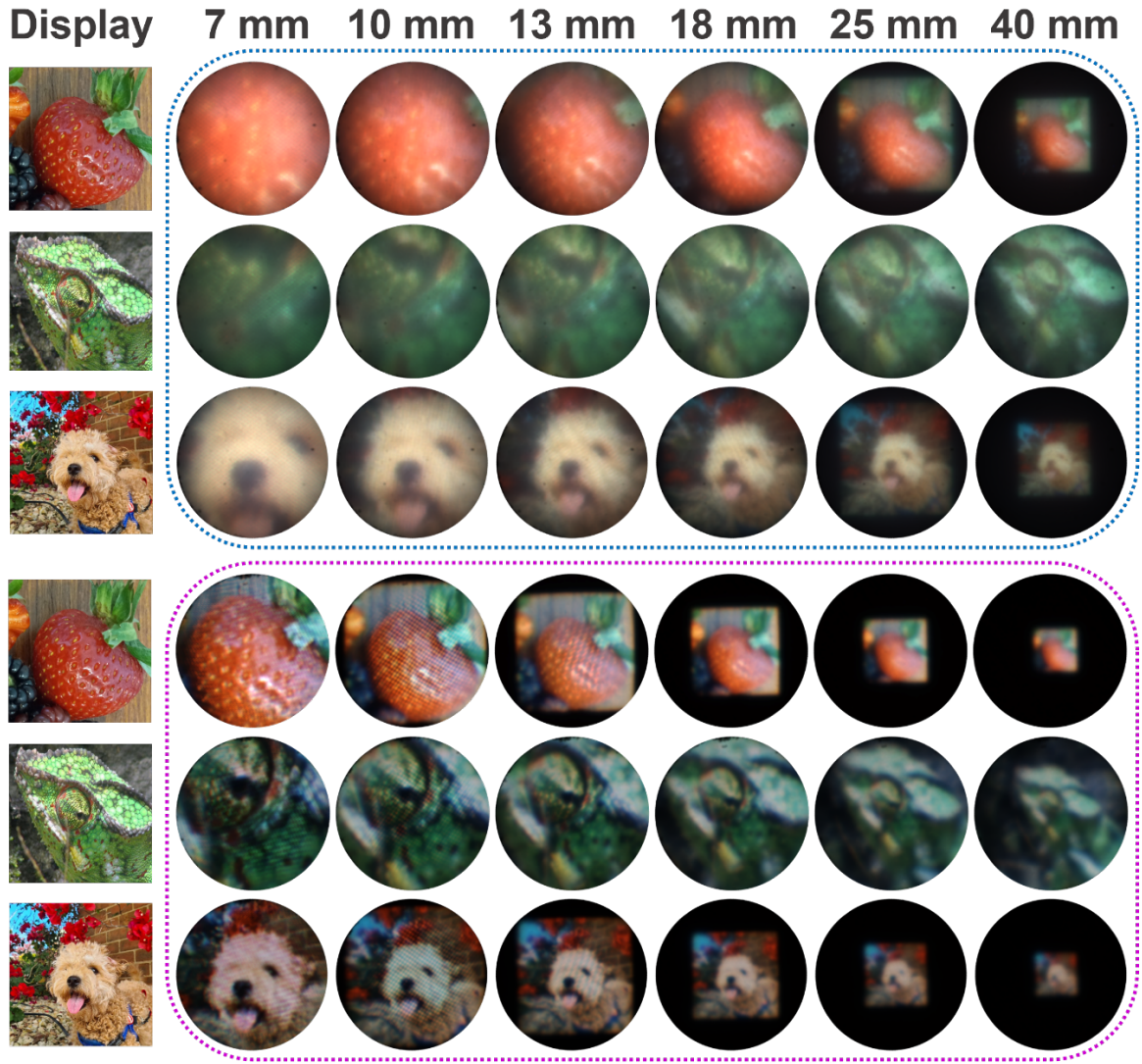

*Figure S7. Comparison of the depth of field for the MO endoscope and the GRIN endoscope at specified object distances. Images captured with the MO endoscope are displayed in the top 3 rows. Images captured with the GRIN endoscope are displayed in the bottom three rows.*

**S.7 Additional Image Processing Comparison**

As described in the main text, no deconvolution steps were applied on the captured images. For presentation, the images were only slightly blurred and color adjusted. This minimizes artefacts such as moiré patterns, due to the capture of individual cores, as can be seen in Figure S8a. This yielded images as shown in Figure S8b. We emphasize that the relative adjustment of different color channels would not require any significant processing time during video rate imaging, whereas the slight deblurring, if required could also be achieved by defocusing the imaging objective.

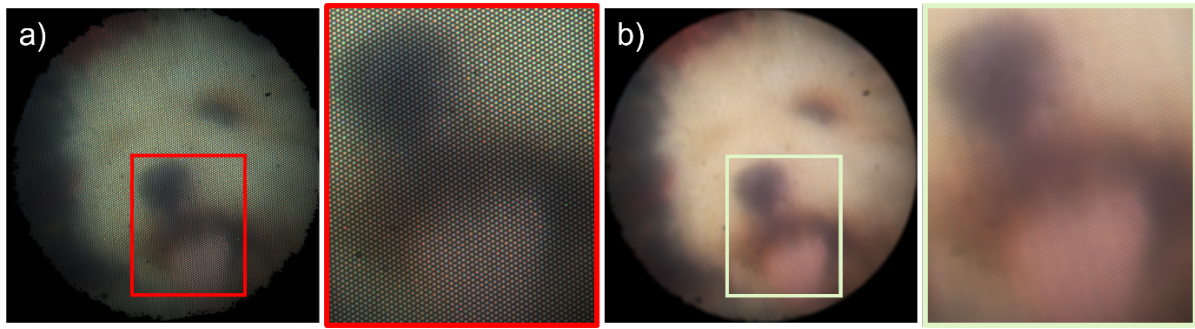

*Figure S8. Example of postprocessing applied to the captured images. a) A captured image and magnified section of that image, showing the appearance of individual fiber cores. b) The same image, after blurring, brightness/contrast, and color channel adjustments.*

# References

[1] Huang, Luocheng, et al. "Full-Color Metaoptical Imaging in Visible Light." *Advanced Photonics Research* 3.5 (2022): 2100265.